# Global Research Trends in the Modern Language Journal from 1999 to 2018: A Data-Driven Analysis

Dr. M Sadik Batcha[1], Younis Rashid Dar[2], Muneer Ahmad[3*]

[1]Mentor and Research Supervisor, Professor and University Librarian, Annamalai University, Annamalai nagar Tamil Nadu

[2]Ph. D Research Fellow, Department of Linguistics, University of Kashmir, Hazratbal Srinagar, Jammu & Kashmir

[3]Ph. D Research Scholar, Department of Library and Information Science, Annamalai University, Annamalai nagar, Tamil Nadu

*Corresponding Author: Muneer Ahmad, Ph. D Research Scholar, Department of Library and Information Science, Annamalai University, Annamalai nagar, Tamil Nadu, Email: muneerbangroo@gmail.com

**ABSTRACT**

*The present study conducts a scientometric study of the Modern Language Journal literature from 1999 to 2018 based on the database of Web of Science, 2018. A total of 2564 items resulted from the publication name using "Modern Language Journal" as the search term. Based on the number of publications during the study period no consistent growth is observed in the research activities pertaining to the journal. The annual distribution of publications, number of authors, institution productivity, country wise publications and Citations are analyzed. Highly productive authors, institutions, and countries are identified. The results reveal that the maximum number of papers 179 is published in the year 1999. It was also observed that Byrnes H is the most productive, contributed 51 publications and Kramsch C is most cited author in the field having 543 global citations. The highest number (38.26%) of publications, contributed from USA and the foremost productive establishment was University of Iowa.*

**Keywords:** *Citations, the Modern Language Journal, Bibexcel, Most Productive Authors, Histcite, VOS viewer, ACPP, Language.*

## INTRODUCTION

The study of language has fascinated researchers since ages yet in many ways we are just at the beginning to understand the complex nature of languages and such a study finds an important place in the overall understanding of the structure of human existence. Language is an important tool of communication and imagining a life without language is extremely difficult. The study of Language/s from a scientific perspective is the subject matter of the field of Linguistics. The field of linguistics has a direct relationship and impact on different areas of investigation as diverse as: sociology, anthropology, philosophy, psychology, artificial intelligence, cognitive neuroscience, among others (Akmajian, Demers, Farmer, & Harnish, 2010).

The aim of this paper is to study the scholarly communications of the Modern Language Journal. The MLJ is an international refereed journal that is dedicated to promoting scholarly exchange among researchers and teachers of all modern foreign languages and English as a second language. The journal is particularly committed to publishing high quality work in non-English languages. The editorial mission of *The Modern Language Journal* is to publish "research and discussion about the learning and teaching of foreign and second languages". Its publication focus is further defined by linking the findings of research to teaching and learning in a variety of settings and on all educational levels. Article contributions are expected to meet the highest standards of scholarly excellence, advance theoretical knowledge, and explore clearly stated and well supported implications for teaching (Loewen, 2019).

## REVIEW OF LITERATURE

There have been enormous amount of scientometric studies all across the world. Some of the relevant studies in the aforesaid direction are worthy of examinations. (Batcha & Ahmad, 2017) analysed comparative analysis of Indian Journal of Information Sources and Services (IJISS) and Pakistan Journal of Library and Information Science (PJLIS) during 2011-2017 and studied various aspects like year wise distribution of papers, authorship pattern & author productivity, degree of collaboration pattern of Co-Authorship , average length of papers, average keywords, etc and found 138 (94.52%) of contributions from IJISS were made by Indian authors and similarly 94 (77.05) of contributions from PJLIS were done by Pakistani authors. Papers by Indian and Pakistani Authors with Foreign Collaboration are minimal (1.37% of articles) and (4.10% of articles) respectively.

(Batcha, Jahina, & Ahmad, 2018) has examined scientometric analysis of the DESIDOC Journal and analyzed the pattern of growth of the research output published in the journal, pattern of authorship, author productivity, and, subjects covered to the papers over the period (2013-2017). It found that 227 papers were published during the period of study (2001-2012). The maximum numbers of articles were collaborative in nature. The subject concentration of the journal noted was Scientometrics. The maximum numbers of articles (65 %) have ranged their thought contents between 6 and 10 pages.

(Ahmad & Batcha, 2019) analyzed research productivity in Journal of Documentation (JDoc) for a period of 30 years between 1989 and 2018. Web of Science database a service from Clarivate Analytics has been used to download citation and source data. Bibexcel and Histcite application software have been used to present the datasets. Analysis part focuses on the parameters like citation impact at local and global level, influential authors and their total output, ranking of contributing institutions and countries. In addition to this scientographical mapping of data is presented through graphs using VOSviewer software mapping technique.

**(Ahmad, Batcha, Wani, Khan, & Jahina, 2017) explored)** scientometric analysis of the Webology Journal. The paper analyses the pattern of growth of the research output published in the journal, pattern of authorship, author productivity, and subjects covered to the papers over the period (2013-2017). It was found that 62 papers were published during the period of study (2013-2017). The maximum numbers of articles were collaborative in nature. The subject concentration of the journal noted was Social Networking/ Web 2.0/Library 2.0 and Sciento-metrics or Bibliometrics. Iranian researchers contributed the maximum number of articles (37.10%). The study applied standard formula and statistical tools to bring out the factual results.

(Ahmad & Batcha, 2019) studied the scholarly communication of Bharathiar University which is one of the vibrant universities in Tamil Nadu. The study find out the impact of research produced, year-wise research output, citation impact at local and global level, prominent authors and their total output, top journals of publications, collaborating countries, most contributing departments and publication trends of the university during 2009 to 2018. The 10 years' publication data of the university indicate that a total of 3440 papers have been published from 2009 to 2018 receiving 38104 citations with h-index as 68. In addition the study used scientographical mapping of data and presented it through graphs using VOSviewer software mapping technique.

(Ahmad, Batcha, & Jahina, 2019) quantitatively identified the research productivity in the area of artificial intelligence at global level over the study period of ten years (2008-2017). The study identified the trends and characteristics of growth and collaboration pattern of artificial intelligence research output. Average growth rate of artificial intelligence per year increases at the rate of 0.862. The multi-authorship pattern in the study is found high and the average number of authors per paper is 3.31. Collaborative Index is noted to be the highest range in the year 2014 with 3.50. Mean CI during the period of study is 3.24. This is also supported by the mean degree of collaboration at the percentage of 0.83 .The mean CC observed is 0.4635. Lotka's Law of authorship productivity is good for application in the field of artificial intelligence literature. The distribution frequency of the authorship follows the exact Lotka's Inverse Law with the exponent á = 2. The modified form of the inverse square law, i.e., Inverse Power Law with á and C parameters as 2.84 and 0.8083 for artificial intelligence literature is applicable and appears to provide a good fit. Relative Growth Rate [Rt(P)] of an article gradually increases from -0.0002 to 1.5405, correspondingly the value of doubling time of the articles Dt (P) decreases from 1.0998 to 0.4499 (2008-2017). At the outset the study reveals the fact that the artificial intelligence literature research study is one of the emerging and blooming fields in the domain of information sciences.

(Batcha, Dar, & Ahmad, 2019) presented a scientometric analysis of the journal titled "Cognition" for a period of 20 years from 1999 to 2018. The study was conducted with an aim to provide a summary of research activity in the journal and characterize its most aspects. The research coverage includes the year wise distribution of articles, authors, institutions, countries and citation analysis of the journal. The analysis showed that 2870 papers were published in journal of Cognition from 1999 to 2018. The study identified top 20 prolific authors, institutions and countries of the journal. Researchers from USA have made the most percentage of contributions.

### OBJECTIVES OF THE STUDY

The main objective of the study is to consider on the mapping of 2564 articles published by *the Modern Language journal* during the period of 1999 – 2018 and the specific objectives are to identify and carry out the following factors:

- To examine the annual publications output of *the Modern Language journal*.
- To estimate publication density through mapping of top 20 authors, institutions and countries based on their number of research papers.
- Find out the top 20 prolific authors, institutions and countries contributed the journal.

### DATA SOURCE AND METHODOLOGY

For the present study, the database of Web of Science (WoS), a product of the Clarivate Analytics is employed to retrieve bibliographic data of literature on the Modern Language Journal from 1999 to 2018 for this study. WoS is adopted because it is recognized as the leading English-language supplier of services providing access to the published information in the multidiscipline fields of science and technology. Bibliographic Data of 2564 research publications from 1999-2018 was considered for this study. The required data were enhanced using different parameters like title, authors, years, countries, and research institutions. The data were analyzed with Histcite and Bibexcel tools. Further, mapping tool such as VOSviewer was used to study the collaboration behaviour and citation network. Histcite calculates the total local citation score (TLCS) and total global citation score (TGCS). TLCS is the number of times a publication cited by other publications in the current data set which means the citation scored among the collection of 2564 publications of the Modern Language Journal. TGCS is the number of times a publication cited by other publications in WoS.

**Table1.** *Details of the Important Points of the Data Sample During 1999 to 2018.*

| S. No. | Details about Sample | Observed Values |
|---|---|---|
| 1 | Duration | 1999-2018 |
| 2 | Collection Span | 20 Years |
| 3 | Total No. of Records | 2564 |
| 4 | Total No. of Authors | 1682 |
| 5 | Frequently Used Words | 2931 |
| 6 | Document Types | 12 |
| 7 | Languages | 3 |
| 8 | Contributing Countries | 39 |
| 9 | Contributing Institutions | 657 |
| 10 | Institutions with Sub Division | 1088 |
| 11 | Total Cited References | 46842 |
| 12 | Total Local Citation Scores | 1604 |
| 13 | Total Global Citation Scores | 21014 |
| 14 | H-Index | 74 |

### DISCUSSION AND RESULT

**Evaluate the Annual Output of Publications**

The data from table 2 and graph 1 reveals that the numbers of research documents published from 1999 to 2018 shows variation in publication of research articles in the Journal. According to the publication output from the table 2 the year wise distribution of research documents, 1999 has the highest number of research documents 179 (6.98%) with 79 (4.93%) of total local citation score and 1293 (6.15%) of total global citation score values and being prominent among the 20 years output and it stood in first rank position. The year 2010 has 173 (6.75%) research documents and it stood in second position with 54 (3.37%) of total local citation score and 1192 (5.67%) of total global citation score were scaled. It is followed by the year 2007 with 168 (6.55 %) of records and it stood in third rank position along with 134 (8.35%) of total local citation score and 2147 (10.22%) of total global citation score measured. The year 2009 has 162 (6.32%) research documents and it stood in fourth position with 114 (7.11%) & 1616 (7.69%) of total global citation score. The year 2003 has 158 (6.16%) research documents and it stood in 5th position with 76 (4.74%) of total local citation score and 1243 (5.92%) of total global citation score were scaled. It has been observed from the data that the increase in publications in the journal is not directly linked to an increase in the overall citation score of the research articles. Graph 1 presents the year wise publications and depicts the citation score. It clearly indicates on the fact that increase in publication rate is not directly linked to increase in citation score.

**Global Research Trends in the Modern Language Journal from 1999 to 2018: A Data-Driven Analysis**

**Table2.** *Annual Distribution of Publications and Citations*

| S. No. | Year | Publications | % | Rank | TLCS | % | Rank | TGCS | % | Rank |
|---|---|---|---|---|---|---|---|---|---|---|
| 1 | 1999 | 179 | 6.98 | 1 | 79 | 4.93 | 9 | 1293 | 6.15 | 7 |
| 2 | 2000 | 118 | 4.60 | 14 | 48 | 2.99 | 17 | 724 | 3.45 | 13 |
| 3 | 2001 | 129 | 5.03 | 11 | 81 | 5.05 | 7 | 1078 | 5.13 | 12 |
| 4 | 2002 | 156 | 6.08 | 6 | 91 | 5.67 | 6 | 1360 | 6.47 | 5 |
| 5 | 2003 | 158 | 6.16 | 5 | 76 | 4.74 | 12 | 1243 | 5.92 | 10 |
| 6 | 2004 | 140 | 5.46 | 9 | 170 | 10.60 | 1 | 1556 | 7.40 | 4 |
| 7 | 2005 | 118 | 4.60 | 14 | 133 | 8.29 | 3 | 1816 | 8.64 | 2 |
| 8 | 2006 | 151 | 5.89 | 7 | 81 | 5.05 | 7 | 1280 | 6.09 | 8 |
| 9 | 2007 | 168 | 6.55 | 3 | 134 | 8.35 | 2 | 2147 | 10.22 | 1 |
| 10 | 2008 | 148 | 5.77 | 8 | 77 | 4.80 | 11 | 1346 | 6.41 | 6 |
| 11 | 2009 | 162 | 6.32 | 4 | 114 | 7.11 | 4 | 1616 | 7.69 | 3 |
| 12 | 2010 | 173 | 6.75 | 2 | 54 | 3.37 | 15 | 1192 | 5.67 | 11 |
| 13 | 2011 | 135 | 5.27 | 10 | 67 | 4.18 | 13 | 1265 | 6.02 | 9 |
| 14 | 2012 | 109 | 4.25 | 17 | 40 | 2.49 | 18 | 542 | 2.58 | 16 |
| 15 | 2013 | 121 | 4.72 | 13 | 79 | 4.93 | 9 | 707 | 3.36 | 14 |
| 16 | 2014 | 116 | 4.52 | 16 | 59 | 3.68 | 14 | 698 | 3.32 | 15 |
| 17 | 2015 | 128 | 4.99 | 12 | 111 | 6.92 | 5 | 438 | 2.08 | 17 |
| 18 | 2016 | 54 | 2.11 | 18 | 53 | 3.30 | 16 | 435 | 2.07 | 18 |
| 19 | 2017 | 51 | 1.99 | 19 | 39 | 2.43 | 19 | 207 | 0.99 | 19 |
| 20 | 2018 | 50 | 1.95 | 20 | 18 | 1.12 | 20 | 71 | 0.34 | 20 |
| | Total | 2564 | 100.00 | | 1604 | 100.00 | | 21014 | 100.00 | |

*\*TLCS = Total Local Citation Score, \*TGCS = Total Global Citation Score*

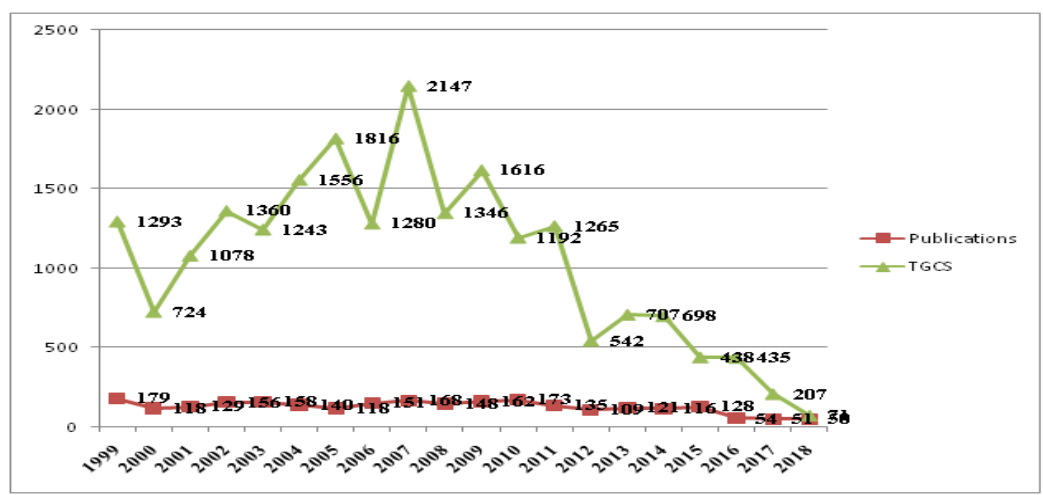

**Graph1.** *Annual Distribution of Publications and Global Citations*

### Analysis of the Publication Output of Top 20 Authors

The ranking of authors of various research articles is displayed in Table 3 and figure 1. In the rank analysis, the authors who have published 9 articles or more are considered into account to avoid a long list. It is observed that there are a total of 1682 authors for 2564 records and it shows the top 20 most productive authors during 1999-2018. Byrnes H published 51 (1.99%) articles with 274 TGCS articles, followed by Magnan SS 22 (0.86%) with 43 TGCS articles, Lafford BA 15 (0.59%) with 79 TGCS articles, McGinnis S 15 (0.59%) with 29 TGCS articles, Call ME 14 (0.55%) with zero TGCS article, Schultz JM 13 (0.51%) with 38 TGCS articles, Benseler DP with 11 articles (0.43%) with 03 TGCS and other authors have contributed comparatively less than the top seven authors during the period of study. The data set clearly indicates that an increase in number of publications of an author has no direct impact in the increase in overall citation score. It is found that the ranked contributors are from the following research Institutions: University of Iowa, Georgetown University, Ohio State University, Penn State University, University of Wisconsin and so on. It could be identified from the author wise analysis, the following authors Byrnes H, Magnan SS, Lafford BA, McGinnis S, Call ME, Schultz JM and Benseler DP were identified the most productive authors based on the number of research papers published in the Journal. The data set puts forth that the authors Kramsch C with 543 citations, Byrnes H with 274 citations, Lantolf JP with 199 citations and Garcia O with 197 citations.

**Table3.** *Publication output of Top 20 Authors and Citation Score*

| S. No. | Authors | Publications | % | TLCS |
|---|---|---|---|---|
| 1 | Byrnes H | 51 | 1.99 | 55 |
| 2 | Magnan SS | 22 | 0.86 | 10 |
| 3 | Lafford BA | 15 | 0.59 | 10 |
| 4 | McGinnis S | 15 | 0.59 | 1 |
| 5 | Call ME | 14 | 0.55 | 0 |
| 6 | Schultz JM | 13 | 0.51 | 5 |
| 7 | Benseler DP | 11 | 0.43 | 1 |
| 8 | Herschensohn J | 11 | 0.43 | 1 |
| 9 | Kramsch C | 11 | 0.43 | 58 |
| 10 | Severino C | 11 | 0.43 | 0 |
| 11 | Garcia O | 10 | 0.39 | 10 |
| 12 | Kaye AS | 10 | 0.39 | 0 |
| 13 | Kinginger C | 10 | 0.39 | 16 |
| 14 | Nassaji H | 10 | 0.39 | 11 |
| 15 | Goebel RO | 9 | 0.35 | 0 |
| 16 | Hall JK | 9 | 0.35 | 34 |
| 17 | Lantolf JP | 9 | 0.35 | 33 |
| 18 | Murphy D | 9 | 0.35 | 4 |
| 19 | Nuessel F | 9 | 0.35 | 0 |
| 20 | Pino BG | 9 | 0.35 | 0 |

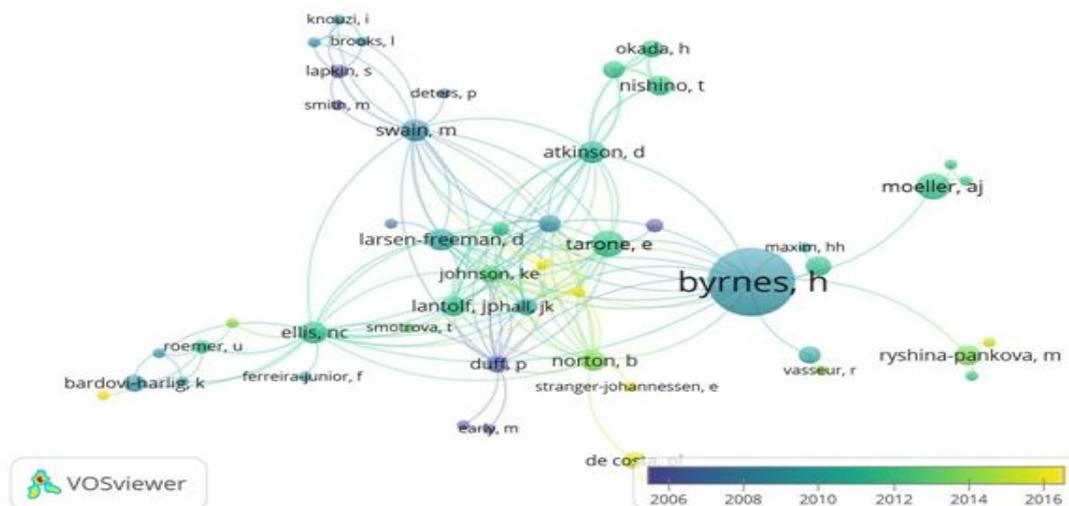

**Figure1.** *Highly Prolific Authors*

## Analysis of the Publication Output of Top 20 Institutions

The individualities of 20 most productive institutions were analyzed in this part. Institutions that published more than 21 and above publications have been considered as highly productive institutions. Table 4 Summarizes articles, the global citation score, local citation score and average author per paper of the publications of these institutions. In total, 657 institutions, including 1088 subdivisions published 2564 research papers during 1999 – 2018. The mean average is 3.90 research articles per Institution. Out of 657 institutions, top 20 institutions published 736 (28.71%) research papers and the rest of the institution published 1828 (71.29%) research papers respectively. Based on the number of published research records the institutions are ranked. The Institutions are ranked on the basis of number of research records published during 1999-2018 as:

The institution "University of Iowa" holds the first rank and the institution published 96 (3.74%) research papers with 06 local and 70 global citation scores, the average citation per paper is 0.73. The second rank is achieved by

**Global Research Trends in the Modern Language Journal from 1999 to 2018: A Data-Driven Analysis**

"Georgetown University" the institution published 71 (2.77%) research papers with 76 local and 646 global citation scores, the average citation per paper is 9.10. The "Ohio State University" holds the 3rd rank, the institution published 61 (2.38%) research papers with 03 local and 42 global citation scores, the average citation per paper is 0.69. The "Penn State University" holds the 4th rank, the institution published 60 (2.34%) research papers with 86 local and 1410 global citation scores, the average citation per paper is 23.50. The "University of Wisconsin" holds the 5th rank, the institution published 47 (1.83%) research papers with 56 local and 412 global citation scores, the average citation per paper is 8.77. It is clear from the analysis that following institutions: University of Iowa, Georgetown University, Ohio State University, Penn State University and University of Wisconsin among others were identified the most productive institutions based on the number of research papers published in the Journal. However, University of California, Berkley (29.91), Penn State University (23.50), Arizona University (20.18), University of Texas (19.08) and University of Pennsylvania (18.00) are the institutions with high ACPP indicating the quality work with high citation impact; hence they can be recognized as the most productive institutions based on the annual citation per paper received in terms of publications.

**Table4.** *Institutional Ranking and Research Performance*

| S. No. | Institution | Publications | % | TLCS | TGCS |
| --- | --- | --- | --- | --- | --- |
| 1 | University of Iowa | 96 | 3.74 | 6 | 70 |
| 2 | Georgetown University | 71 | 2.77 | 76 | 646 |
| 3 | Ohio State University | 61 | 2.38 | 3 | 42 |
| 4 | Penn State University | 60 | 2.34 | 86 | 1410 |
| 5 | University of Wisconsin | 47 | 1.83 | 56 | 412 |
| 6 | University of Minnesota | 43 | 1.68 | 63 | 570 |
| 7 | Brigham Young University | 34 | 1.33 | 1 | 82 |
| 8 | Arizona State University | 31 | 1.21 | 19 | 235 |
| 9 | Georgia State University | 29 | 1.13 | 13 | 355 |
| 10 | Michigan State University | 28 | 1.09 | 21 | 313 |
| 11 | University of Maryland | 27 | 1.05 | 13 | 268 |
| 12 | University of Illinois | 26 | 1.01 | 44 | 337 |
| 13 | University of Arizona | 25 | 0.98 | 46 | 398 |
| 14 | University of Texas | 25 | 0.98 | 20 | 477 |
| 15 | Indiana University | 23 | 0.90 | 23 | 136 |
| 16 | University of California, Berkeley | 23 | 0.90 | 68 | 688 |
| 17 | University of Pennsylvania | 23 | 0.90 | 47 | 414 |
| 18 | Arizona University | 22 | 0.86 | 12 | 444 |
| 19 | Carnegie Mellon University | 21 | 0.82 | 7 | 155 |
| 20 | Iowa State University | 21 | 0.82 | 9 | 107 |

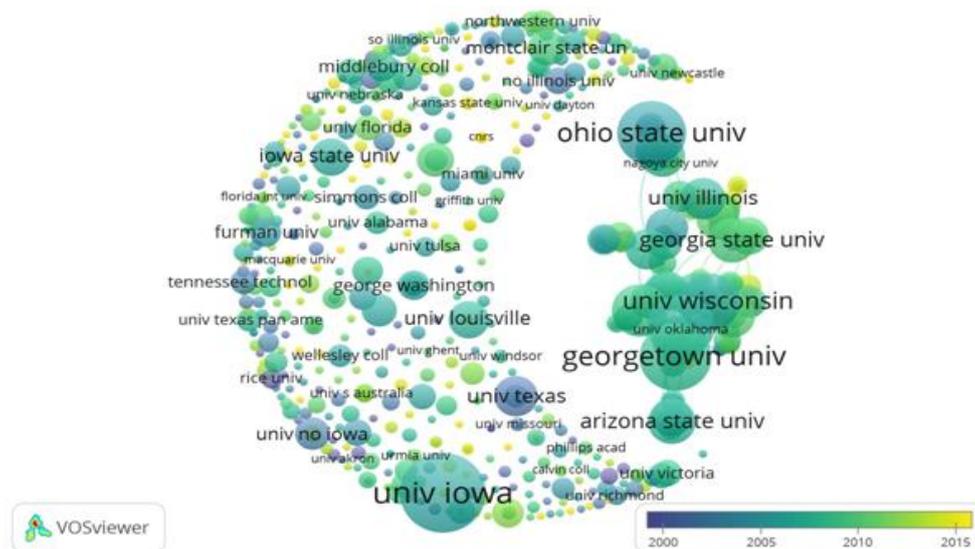

**Figure2.** *Collaboration of Institutions and their clusters*

# Global Research Trends in the Modern Language Journal from 1999 to 2018: A Data-Driven Analysis

## Analysis of the Publication Output of Top 20 Countries

Table 5 and Figure 3 displays the publication output of the top twenty countries by number of papers and USA acquired 1st rank among the top twenty countries under consideration with its total global citation score 10410. Among all 39 countries that participated in research during 1999 and 2018, the countries that rank between 2[nd] and 20[th] position are: Canada, UK, Japan, Australia, China, Netherlands, Spain, New Zealand, Sweden, Germany, South Korea, Iran, Denmark, Finland, Israel, Belgium, Taiwan, France and Ireland. By using Country Mapping Analysis, it has been found that the nodes are linked to each other indicating that countries are having collaboration with other associated nations. It could be identified from the analysis the following countries: USA, Canada, UK, Japan, Australia, Peoples Republic of China and Netherlands were identified the most productive countries based on the number of research papers published.

**Table 5.** *Distribution of the Publication Output of Top 20 Countries*

| S. No. | Country | Publications | % | TLCS | TGCS |
|---|---|---|---|---|---|
| 1 | USA | 981 | 38.26 | 803 | 10410 |
| 2 | Canada | 95 | 3.71 | 129 | 2069 |
| 3 | UK | 76 | 2.96 | 195 | 2388 |
| 4 | Japan | 39 | 1.52 | 90 | 1058 |
| 5 | Australia | 31 | 1.21 | 54 | 820 |
| 6 | Peoples R China | 23 | 0.90 | 16 | 488 |
| 7 | Netherlands | 21 | 0.82 | 29 | 484 |
| 8 | Spain | 20 | 0.78 | 25 | 437 |
| 9 | New Zealand | 14 | 0.55 | 52 | 394 |
| 10 | Sweden | 14 | 0.55 | 30 | 136 |
| 11 | Germany | 12 | 0.47 | 17 | 152 |
| 12 | South Korea | 10 | 0.39 | 13 | 119 |
| 13 | Iran | 9 | 0.35 | 5 | 121 |
| 14 | Denmark | 8 | 0.31 | 28 | 245 |
| 15 | Finland | 7 | 0.27 | 3 | 77 |
| 16 | Israel | 7 | 0.27 | 6 | 111 |
| 17 | Belgium | 6 | 0.23 | 3 | 51 |
| 18 | Taiwan | 6 | 0.23 | 5 | 139 |
| 19 | France | 5 | 0.20 | 31 | 193 |
| 20 | Ireland | 3 | 0.12 | 0 | 10 |

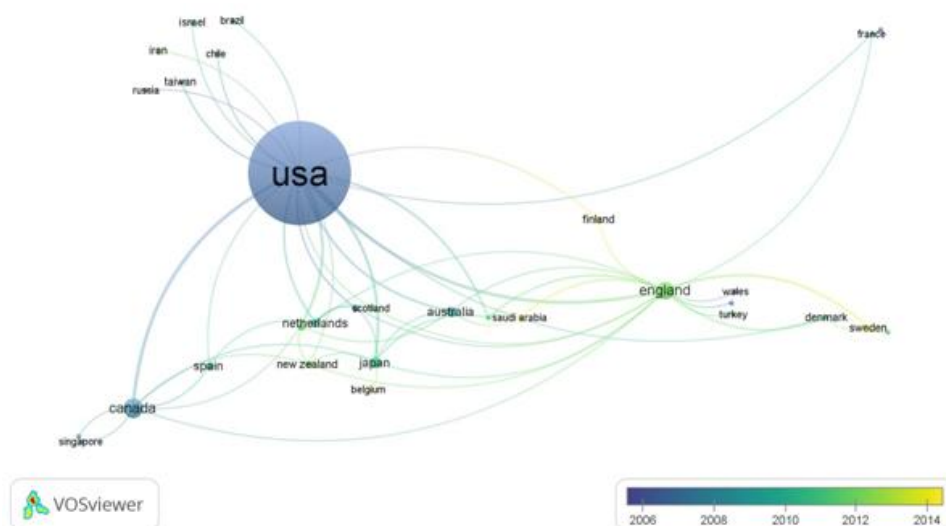

**Figure 3.** *Ranking of Country wise Distribution*

## CONCLUSION

The number of papers published in the Modern Language Journal has gradually decreased during 1999–2018 and the study has shown that a total number of 2564 research documents have been published during a period of 20 years. The data from this paper suggest that the authors Byrnes H, Magnan SS, Lafford BA, McGinnis S, Call ME, Schultz JM and Benseler DP were identified as the most prolific authors based on the number of research papers contributed. It could be seen from the institutions wise analysis that the following institutions: University of Iowa, Georgetown University, Ohio State University, Penn State University and University of Wisconsin have published maximum number of research papers in the journal. The following countries: USA Canada, UK, Japan, Australia, China, Netherlands, Spain, New Zealand and Sweden were recognized the nations that have contributed highest number of publications during the period under study.

## REFERENCES

[1] Ahmad, M., & Batcha, M. S. (2019a). Mapping of Publications Productivity on Journal of Documentation 1989-2018 : A Study Based on Clarivate Analytics – Web of Science Database. *Library Philosophy and Practice (E-Journal)*, 22 13–2226. Retrieved from https://digitalcommons.unl.edu/libphilprac/2213/

[2] Ahmad, M., & Batcha, M. S. (2019b). Scholarly Communications of Bharathiar University on Web of Science in Global Perspective : A Scientometric Assessment. *Research Journal of Library and Information Science*, *3*(3), 22–29.

[3] Ahmad, M., Batcha, M. S., & Jahina, S. R. (2019). Testing Lotka's Law and Pattern of Author Productivity in the Scholarly Publications of Artificial Intelligence. *Library Philosophy and Practice (E-Journal)*. Retrieved from https://digitalcommons.unl.edu/libphilprac/2716

[4] Ahmad, M., Batcha, M. S., Wani, B. A., Khan, M. I., & Jahina, S. R. (2017). Research Output of Webology Journal ( 2013-2017 ): A Scientometric Analysis. *International Journal of Movement Education and Social Science*, *7*(3), 46–58.

[5] Akmajian, A., Demers, R. A., Farmer, A. K., & Harnish, R. M. (2010). *Linguistics: An Introduction to Language and Communication* (6th ed.). The MIT Press.

[6] Batcha, M. S., Dar, Y. R., & Ahmad, M. (2019). Impact and Relevance of Cognition Journal in the Field of Cognitive Science: An Evaluation. *Research Journal of Library and Information Science*, *3*(4), 21–28.

[7] Batcha, M. S., Jahina, S. R., & Ahmad, M. (2018). Publication Trend in DESIDOC Journal of Library and Information Technology during 2013-2017 : A Scientometric Approach. *International Journal of Research in Engineering, IT and Social Sciences*, *8*(4), 76–82.

[8] Batcha, M. S. & Ahmad, M. (2017). Publication Trend in an Indian Journal and a Pakistan Journal : A Comparative Analysis using Scientometric Approach. *Journal of Advances in Library and Information Science*, *6*(4), 442–449.

[9] Loewen, S. (2019). The Modern Language Journal. Retrieved December 1, 2019, from https://onlinelibrary.wiley.com/journal/15404781